\documentclass[aps, prd, amsmath, floats, floatfix, twocolumn, nofootinbib, superscriptaddress]{revtex4-1}
\usepackage[pdftex]{graphicx}
\usepackage{color}
\usepackage{latexsym}
\usepackage[colorlinks]{hyperref}

\newcommand{\be}{\begin{eqnarray}}
\newcommand{\ee}{\end{eqnarray}}

\begin{document}

\title{Testing the Keplerian disk hypothesis using X-ray reflection spectroscopy}

\author{Ashutosh~Tripathi}
\affiliation{Center for Field Theory and Particle Physics and Department of Physics, Fudan University, 200438 Shanghai, China}

\author{Biao~Zhou}
\affiliation{Center for Field Theory and Particle Physics and Department of Physics, Fudan University, 200438 Shanghai, China}

\author{Askar~B.~Abdikamalov}
\affiliation{Center for Field Theory and Particle Physics and Department of Physics, Fudan University, 200438 Shanghai, China}
\affiliation{Ulugh Beg Astronomical Institute, Tashkent 100052, Uzbekistan}

\author{Dimitry~Ayzenberg}
\affiliation{Center for Field Theory and Particle Physics and Department of Physics, Fudan University, 200438 Shanghai, China}
\affiliation{Theoretical Astrophysics, Eberhard-Karls Universit\"at T\"ubingen, 72076 T\"ubingen, Germany}

\author{Cosimo~Bambi}
\email[Corresponding author: ]{bambi@fudan.edu.cn}
\affiliation{Center for Field Theory and Particle Physics and Department of Physics, Fudan University, 200438 Shanghai, China}

\author{Sourabh~Nampalliwar}
\affiliation{Theoretical Astrophysics, Eberhard-Karls Universit\"at T\"ubingen, 72076 T\"ubingen, Germany}

\begin{abstract}
The Novikov-Thorne model is the standard framework for the description of geometrically thin and optically thick accretion disks around black holes and is widely used to study the electromagnetic spectra of accreting black holes. One of the assumptions of the model is that the particles of the gas move on nearly-geodesic circular orbits on the equatorial plane. In this work, we propose to test the Keplerian velocity of the particles in the accretion disk using X-ray reflection spectroscopy. We present a modified version of {\sc relxill} in which we introduce a phenomenological parameter, $\alpha$, to quantify possible deviations from Keplerian motion. We use our model to fit a \textsl{Suzaku} observation of the black hole binary GRS~1915+105. We find that the estimate of $\alpha$ is correlated to that of the inclination angle of the disk, $i$, and that we could test the Keplerian disk hypothesis in the presence of a robust and independent measurement of $i$. 
\end{abstract}

\maketitle

%%%%%%%%%%%%%%%%%%%%%%%%%%%%%%%

\section{Introduction}

The study of the properties of the electromagnetic radiation emitted from the inner part of the accretion disks around black holes is today an active line of research to investigate the physics and the astrophysics of these objects~\cite{r1,r2,r3,r4}. The interpretation of observational data always requires theoretical models and, as in any other astrophysical system, we can measure the properties of an accreting black hole by fitting its observational data with a proper theoretical model.

The Novikov-Thorne model~\cite{ntm,ntm2} is the most popular accretion disk model to describe the accretion flow around stellar-mass black holes in X-ray binary systems and supermassive black holes in active galactic nuclei (AGNs). The model requires that the spacetime is stationary, axisymmetric, and asymptotically flat. The disk is assumed to be on the equatorial plane, and the particles of the disk move on nearly-geodesic, circular orbits (Keplerian motion). The radial heat transport is supposed to be negligible compared to the energy radiated from the disk's surface. The conservation laws for rest-mass, energy, and angular momentum impose a set of equations for the time-averaged radial structure of the disk~\cite{ntm2}. The inner edge of the disk is often assumed to be at the innermost stable circular orbit (ISCO). When the gas reaches the ISCO, it quickly plunges onto the black hole without an appreciable emission of additional radiation.

Understanding the validity and the limitations of the Novikov-Thorne model is crucial for a correct interpretation of the observational data and reliable measurements of the properties of accreting black holes. Indeed, if some of the model assumptions were not satisfied by the source that we want to study, the final measurements would be affected by undesirable systematic uncertainties. For example, black hole spin measurements with the continuum-fitting method~\cite{cfm1,cfm2} or X-ray reflection spectroscopy~\cite{ref1,ref2} normally employ the Novikov-Thorne model and therefore any deviations from it could have an impact on the inferred spin values.

There are already a number of publications in literature studying the validity and the limitations of the Novikov-Thorne model. Generally speaking, the model is thought to describe well the accretion flow around a black hole when the source is in the thermal state and its Eddington-scaled accretion disk luminosity is between $\sim 5$\% to $\sim 30$\%~\cite{mcc06,pen10,ste10}, even if such a condition is not rarely forgotten in observational studies, in part because it is difficult to verify and in part because of the lack of alternative theoretical models. Systematic uncertainties on black hole spin measurements arising from the use of the Novikov-Thorne model were investigated through numerical simulations for the continuum-fitting method in Ref.~\cite{kul11} and X-ray reflection spectroscopy in Ref.~\cite{rey08}. The impact of the radiation from the plunging region was studied in Refs.~\cite{zhu12,aca,ml}. The role of the thickness of the disk was investigated in Refs.~\cite{taylor,taylor2,ml2,shafqat1,shafqat2}.

In the present work, we propose to test the assumption that the particles of the disk move on nearly-geodesic circular orbits (Keplerian disk hypothesis) using X-ray reflection spectroscopy. Magnetic fields or viscous stress may indeed induce deviations from geodesic motion of the disk's particles. We modify the relativistic reflection model {\sc relxill}~\cite{jt1,jt2,jt3} by introducing a phenomenological parameters, $\alpha$, to parametrize the angular velocity of the gas in the accretion disk. For $\alpha = 0$, we have a Keplerian disk. For $\alpha > 0$ ($< 0$), the accretion disk has a higher (lower) angular velocity and we have a super-Keplerian (sub-Keplerian) disk. $\alpha$ can be a free parameter in the model. From the comparison of the theoretical predictions with X-ray data of specific sources, we can constrain the value of $\alpha$ and check whether it is consistent with zero, namely we can confirm the Keplerian disk hypothesis.

The content of the paper is as follows. In Section~\ref{s-model}, we present our phenomenological model with the parameter $\alpha$ to test the Keplerian disk hypothesis. In Section~\ref{s-grs}, we apply our model to a \textsl{Suzaku} observation of the stellar-mass black hole in the X-ray binary GRS~1915+105 in order to illustrate the possibility of testing the Keplerian disk hypothesis with our method. We discuss our results in Section~\ref{s-dis}.

\section{A model to test the Keplerian disk hypothesis \label{s-model}}

Relativistic reflection features are commonly observed in the X-ray spectra of black hole binaries and AGNs~\cite{tanaka95,nandra07,blum09,fabian12,walton13,cao18,tripathi19}. They are thought to be generated by illumination of a cold accretion disk ($T_{\rm disk} \sim 0.1$-1~keV for stellar-mass black holes and $T_{\rm disk} \sim 1$-10~eV for supermassive black holes) by a corona~\cite{galeev79,st79,gf91}, which is some hot electron cloud ($T_{\rm e} \sim 100$~keV) near the black hole. Thermal photons from the accretion disk inverse Compton scatter off free electrons in the corona and this produces a power law spectrum ($dN_\gamma/dE \propto E^{-\Gamma}$, where $\Gamma$ is called the photon index) with a high energy exponential cutoff ($E_{\rm cut} \approx 2$-3~$T_{\rm e}$). A fraction of the Comptonized photons illuminate the disk, producing the reflection spectrum. The most prominent features in the reflection spectrum are usually the iron K$\alpha$ complex in the soft X-ray band and the Compton hump peaked at 20-30~keV~\cite{reflionx,gk10}.

Relativistic reflection spectra of accretion disks are normally calculated in two steps. First, we calculate the non-relativistic reflection spectrum at the emission point in the rest-frame of the gas. These calculations involve only atomic physics. Second, we convolve the reflection spectrum at any point on the disk to get the spectrum of the whole disk as observed far from the source. Here a ray-tracing code calculates the photons trajectories, from the emission point on the disk to the detection point at the location of the distant observer, and we employ a disk model to calculate the effect of Doppler boosting on the reflection spectrum. These calculations have been extensively discussed in literature; see, e.g., Refs.~\cite{nk1,nk2}.

The flux of the radiation emitted by the accretion disk and detected by a distant observer can be written as~\cite{nk1}
\be\label{eq-flux}
F (\nu_{\rm o}) &=& \frac{1}{D^2} \int I_{\rm o} (\nu_{\rm o},X,Y) \, dX dY \nonumber\\
&=& \frac{1}{D^2} \int g^3 I_{\rm e} (\nu_{\rm e},r_{\rm e},\vartheta_{\rm e}) \, dX dY \, ,
\ee
where $\nu_{\rm o}$ and $\nu_{\rm e}$ are, respectively, the photon frequencies at the detection point of the observer and at the emission point in the rest-frame of the gas, $D$ is the distance of the source from the observer, and $I_{\rm o}$ and $I_{\rm e}$ are, respectively, the specific intensity of the radiation measured by the observer and at the emission point in the rest-frame of the gas. $r_{\rm e}$ is the emission radius on the disk and $\vartheta_{\rm e}$ is the emission angle, namely the angle between the disk normal and the photon propagation direction in the rest-frame of the gas. $X$ and $Y$ are the Cartesian coordinates of the image plane of the observer. $g = \nu_{\rm o}/\nu_{\rm e}$ is the redshift factor and $I_{\rm o} = g^3 I_{\rm e}$ follows from Liouville's theorem~\cite{lindquist66}. The calculation of the photon trajectories from the disk to the detection point connects a point on the disk with a point with coordinates $(X,Y)$ on the plane of the observer.

In the Novikov-Thorne model, the particles in the disk move on equatorial, nearly-geodesic, circular orbits. We require that the spacetime is stationary, axisymmetric, and asymptotically flat, and that the line element can be written as
\be
ds^2 &=& g_{tt} dt^2 + 2 g_{t\phi} dt d\phi + g_{rr} dr^2 \nonumber\\
&& + g_{\theta\theta} d\theta^2 + g_{\phi\phi} d\phi^2 \, ,
\ee
where the metric coefficients are independent of the $t$ and $\phi$ coordinates. Geodesic trajectories can be derived from the Lagrangian
\be
\mathcal{L} = \frac{1}{2} g_{\mu\nu} \dot{x}^\mu \dot{x}^\nu \, .
\ee
If we write the Euler-Lagrange equation for the $r$ coordinate and we impose that the motion is on the equatorial plane ($\dot{\theta} = 0$) and on circular orbits ($\dot{r}=\ddot{r}=0$), we find
\be\label{eq-geo}
\left( \partial_r g_{tt} \right) \dot{t}^2 + 2 \left( \partial_r g_{t\phi} \right) \dot{t} \dot{\phi} + \left( \partial_r g_{\phi\phi} \right) \phi^2 = 0 \, .
\ee
The Keplerian angular velocity of the gas in the accretion disk is $\Omega_{\rm K} = \dot{\phi}/\dot{t}$ and, from Eq.~(\ref{eq-geo}), we find the following expression
\be
\Omega_{\rm K} = \frac{-\partial_r g_{t\phi} \pm \sqrt{\left( \partial_r g_{t\phi} \right)^2 - \left( \partial_r g_{tt} \right)\left( \partial_r g_{\phi\phi} \right)}}{\partial_r g_{\phi\phi}} \, .
\ee

If the particles in the accretion disk do not move on geodesic trajectories, we can write their angular velocity as
\be\label{eq-alpha}
\Omega = \left(1 + \alpha\right) \Omega_{\rm K} \, ,
\ee
where $\alpha$ is a parameter to quantify the deviation from Keplerian motion. In general, we should expect that $\alpha$ depends on the radial coordinate; i.e. $\alpha = \alpha (r)$. Moreover, it would be natural to expect $\alpha \le 0$, so that any mechanism can only slow down the motion of the particle in the accretion disk. However, in this work we will consider the simplest case of $\alpha = {\rm constant}$ and we will permit either positive and negative values of $\alpha$. The upper limit on $\alpha$ is obtained by imposing that the particle velocity never exceeds the speed of light. The lower limit is found requiring that the particles always moves on time-like trajectories\footnote{For fast-rotating black holes, the inner edge of the accretion disk enters the ergoregion, where everything must corotate with the black hole. For sufficiently negative values of $\alpha$, this does not happen on time-like geodesics.}

The 4-velocity of the particles in the accretion disk can be written as
\be\label{eq-4v}
u^\mu = \left( \dot{t} , 0 , 0 , \dot{\phi} \right) = \dot{t} \left( 1 , 0 , 0 , \Omega \right) \, ,
\ee
which reduces to the 4-velocity of the particles in the Novikov-Thorne model for $\Omega = \Omega_{\rm K}$. From the normalization condition $g_{\mu\nu} u^\mu u^\nu = - 1$, we can write $\dot{t}$
\be\label{eq-tdot}
\dot{t} = \frac{1}{\sqrt{-g_{tt} - 2 \Omega g_{t\phi} - \Omega^2 g_{\phi\phi}}} \, .
\ee
The redshift factor $g$ is
\be
g = \frac{-u_{\rm o}^\mu k_\mu}{-u_{\rm e}^\nu k_\nu} \, ,
\ee
where $u^\mu_{\rm o} = (1,0,0,0)$ is the 4-velocity of the distant observer, $u_{\rm e}^\nu$ is the 4-velocity of the gas in the accretion disk given by Eq.~(\ref{eq-4v}), and $k^\mu$ is the photon 4-momentum. From Eq.~(\ref{eq-4v}) and Eq.~(\ref{eq-tdot}), we can write the redshift factor $g$ as
\be\label{eq-g}
g = \frac{\sqrt{-g_{tt} - 2 \Omega g_{t\phi} - \Omega^2 g_{\phi\phi}}}{1 - \lambda\Omega} \, ,
\ee
where $\lambda = k_\phi/k_t$ is a constant of motion along the photon trajectory.

The normal to the disk is 
\be
n^\mu =  \left( 0 , 0 , \sqrt{g^{\theta\theta}} , 0 \right)_{r_{\rm e}, \theta_{\rm e}=\pi/2} = \left( 0 , 0 , \frac{1}{r_{\rm e}} , 0 \right) \, ,
\ee
and can be used to calculate the cosine of the emission angle $\vartheta_{\rm e}$ as
\be\label{eq-ea}
\cos\vartheta_{\rm e} = \frac{n^\mu k_\mu}{u_{\rm e}^\nu k_\nu} = \frac{q g}{r_{\rm e}} \, ,
\ee
where $q^2 = \mathcal{Q}/E^2$ and $\mathcal{Q}$ and $E$ are, respectively, the Carter constant and the energy of the photon.

In the end, the angular velocity of the disk, $\Omega$, appears only in the redshift factor $g$ as shown in Eq.~(\ref{eq-g}). The redshift factor $g$ appears in Eq.~(\ref{eq-flux}) and in the calculation of the emission angle $\vartheta_{\rm e}$, as shown in Eq.~(\ref{eq-ea}). In general, we may expect that a different angular velocity of the accretion disk changes the structure of the whole disk. For example, the inner edge of the disk may not be at the ISCO any longer. However, such effects have a very weak impact for what follows.

To calculate the observed flux in Eq.~(\ref{eq-flux}) of the reflection spectrum of the disk, we modify the model {\sc relxill}~\cite{jt1,jt2,jt3}. The procedure is very similar to that used to test the spacetime metric and the details can be found in previous publications~\cite{nk1,nk2,nk3,nk4}. The atomic physics calculations to infer the reflection spectrum at the emission point in the rest-frame of the gas, $I_{\rm e}$, are not affected by the angular velocity of the disk\footnote{If the iron K$\alpha$ line is, for example, at 6.4~keV in the rest-frame of the gas, this is determined by the atomic energy levels. The angular velocity of the disk affects the line energy measured by the distant observer, and this is taken into account by the redshift factor $g$ when we write $I_{\rm o} = g^3 I_{\rm e}$, where the angular velocity appears as shown in Eq.~(\ref{eq-g}).}, and modifications are only present in the convolution model to calculate the reflection spectrum of the whole disk as detected by the distant observer. Within the formalism of the transfer function~\cite{cunn}, we recalculate the transfer function for a grid of different values of the black hole spin parameter $a_*$, the viewing angle of the disk $i$, and the new parameter $\alpha$ used to test the Keplerian angular velocity of the disk.

\begin{figure*}[t]
\begin{center}
\includegraphics[width=15.0cm,trim={3.0cm 1.5cm 5.5cm 19.5cm},clip]{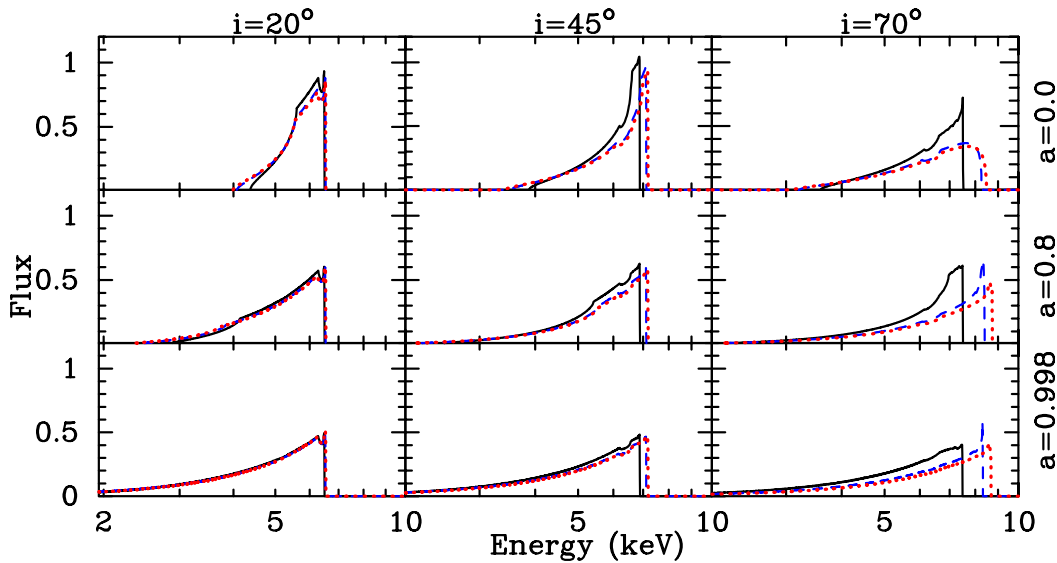}
\end{center}
\vspace{-0.6cm}
\caption{Examples of iron line profiles for a viewing angle $i = 20^\circ$, $45^\circ$, and $70^\circ$ and a black hole spin parameter $a_* = 0$, 0.8, and 0.998. The iron lines profiles for Keplerian disks $\alpha = 0$ (blue dashed profiles) are compared with those from sub-Keplerian disks with $\alpha = -0.2$ (black solid profiles) and super-Keplerian disks with $\alpha = 0.05$ (red dotted profiles). The values $\alpha = -0.2$ and $\alpha = 0.05$ correspond, respectively, to (approximately) the minimum and maximum values for the case $a_* = 0.998$ (see footnote~1). \label{f-lines}}
\end{figure*}

Fig.~\ref{f-lines} shows the impact of the parameter $\alpha$ on a single iron line. This is equivalent to assume that the reflection spectrum at the emission point in the rest-frame of the gas in the disk is only a narrow line at 6.4~keV, but it can help to understand how $\alpha$ affects the redshift factor $g$. The whole reflection spectrum has indeed too many features and the exact shape changes with the ionization parameter $\xi$, so it would not be straightforward to see the role of the parameter $\alpha$. From Fig.~\ref{f-lines}, we see that the iron line profiles are sensitive to the value of $\alpha$ for high viewing angles only. This is perfectly understandable, because $\alpha$ only alters the Doppler boosting and the latter vanishes for $i = 0^\circ$.

\begin{figure}[t]
\begin{center}
\includegraphics[width=8.0cm,trim={2.0cm 0.5cm 3.0cm 18.0cm},clip]{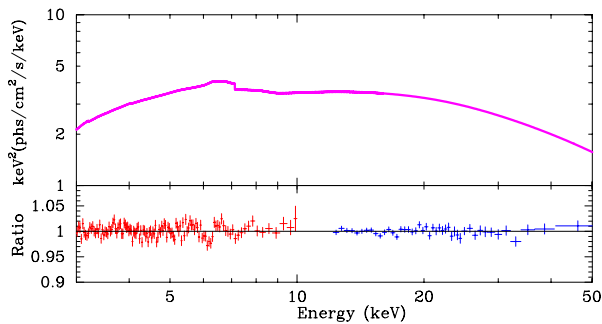}
\end{center}
\vspace{-0.6cm}
\caption{Best-fit model (top panel) and data to best-fit model ratio (bottom panel) for Model~1 with free inclination angle. Red crosses are for XIS1 data and blue crosses are for HXD/PIN data. See the text for more details. \label{f-mr}}
\end{figure}

\section{Analysis of GRS~1915+105 \label{s-grs}}

In this section, we apply our modified version of {\sc relxill} to a specific source to illustrate the capability of our proposal of testing the Keplerian disk hypothesis. We will not enter into technical details related to the data reduction and analysis, which can be found in previous work in literature.

We consider the \textsl{Suzaku} observation of the black hole binary GRS~1915+105 with obs. ID~402071010~\cite{blum09}. The observation was on 2007 May 7 while the source was in the low/hard state. The total exposure time was about 117~ks. We chose this particular observation because it has a number of properties that make it suitable to test extensions of relativistic reflection models~\cite{tlb,nk3,jiachen2}. Here, we also need a source with a high inclination angle of the disk to maximize the impact of $\alpha$, as seen in Fig.~\ref{f-lines}. \textsl{Suzaku} provides a good energy resolution near the iron line in the soft X-ray band as well as data in the hard X-ray band to observe the Compton hump, both quite useful when we study the reflection spectrum of the accretion disk of a black hole. GRS~1915+105 is normally quite a variable and complicated source~\cite{belloni00}, but was instead stable during this 2007 \textsl{Suzaku} observation~\cite{yuexin,tlb}. It showed a simple spectrum with strong relativistic reflection features and no evidence of any disk thermal component~\cite{blum09,yuexin,tlb}. 

We follow the data reduction of Refs.~\cite{yuexin,tlb}. We only use the data from the XIS1 instrument (0.2-12~keV) and from the HXD/PIN instrument (10-70~keV). Eventually we have 2.43~million counts in the XIS1 data (after removing the central region because of pile-up) and 1.36~million counts in the HXD/PIN data.  Data are rebinned to have a minimum of 25~counts per bin in order to use the $\chi^2$ statistics.

We use XSPEC v12.10.1f~\cite{arnaud}. From previous analyses of this observation~\cite{blum09,yuexin,tlb}, we know that the data can be fit well with an absorbed coronal spectrum and disk's reflection component; in XSPEC language, the model is 
\be
\text{\sc tbabs$\times$relxill} \, ,
\ee
where {\sc tbabs} describes the Galactic absorption~\cite{wilms00} and we leave the column density $N_{\rm H}$ free in the fit. {\sc relxill} describes the spectrum from the corona and the reflection spectrum from the accretion disk~\cite{jt1,jt2,jt3}. The former is modeled by a power law with an exponential high energy cutoff and has two free parameters: the photon index $\Gamma$ and the high energy cutoff $E_{\rm cut}$. The reflection spectrum of the disk has eight free parameters. The emissivity profile of the reflection spectrum is modeled by a broken power law and we have thus three free parameters: the inner emissivity index $q_{\rm in}$, the other emissivity index $q_{\rm out}$, and the breaking radius $R_{\rm br}$. The other five parameters are the viewing angle $i$ (i.e. the angle between the black hole spin and the line of sight of the distant observer), the dimensionless black hole spin parameter $a_*$, the ionization parameter of the disk $\xi$, the iron abundance $A_{\rm Fe}$ (in units of the solar iron abundance), and the parameter $\alpha$ in Eq.~(\ref{eq-alpha}) to test the Keplerian motion of the disk. {\sc relxill} has also the normalization of the reflection spectrum and the reflection fraction $R_{\rm f}$ to describe the strength of the reflection component with respect to the power law component from the corona; both parameters are free in the fit. We note that the inner edge of the disk is frozen at the ISCO radius in the fit, but the data require an inner edge very close to the black hole, so we get a very similar result even if we leave the inner edge as a free parameter.

\begin{table*}
\centering
%\label{t-fit}
\scalebox{1.0}{
\begin{tabular}{lccccc}
\hline\hline
& \hspace{0.7cm} Model~1 \hspace{0.7cm} & \hspace{0.7cm} Model~2$a$ \hspace{0.7cm} & \hspace{0.7cm} Model~2$b$ \hspace{0.7cm} & \hspace{0.7cm} Model~2$c$ \hspace{0.7cm} & \hspace{0.7cm} Model~3 \hspace{0.7cm} \\
\hline
{\sc tbabs} && \\
$N_{\rm H} / 10^{22}$ cm$^{-2}$ & $8.35_{-0.07}^{+0.08}$ & $7.69_{-0.07}^{+0.09}$ & $7.88_{-0.07}^{+0.05}$ & $8.09_{-0.07}^{+0.03}$ & $7.16^\star$ \\
\hline
{\sc relxill} && \\
$q_{\rm in}$ & $10.0_{-0.4}^{}$ & $7.4_{-0.9}^{+2.1}$ & $7.4_{-1.0}^{+0.9}$ & $7.4_{-0.8}^{+0.5}$ & $5_{-4}^{+5}$ \\
$q_{\rm out}$ & $0.0_{}^{+0.3}$ & $0.0_{}^{+0.4}$ & $0.0_{}^{+0.3}$ & $0.00_{}^{+0.13}$ & $1.0_{-0.3}^{+0.2}$ \\
$R_{\rm br}$ [M] & $5.10_{-0.30}^{+0.15}$ & $12_{-3}^{+3}$ & $10.9_{-1.9}^{+2.8}$ &  $10.2_{-1.5}^{+1.3}$ & $4_{-3}^{+5}$ \\
$i$ [deg] & $85.2_{-1.0}^{+0.8}$ & $60^\star$ & $65^\star$ & $70^\star$ & $83.6_{-0.4}^{+0.6}$ \\
$a_*$ & $0.9946_{-0.0049}^{+0.0012}$ & $0.957_{-0.005}^{+0.005}$ & $0.976_{-0.003}^{+0.003}$ &  $0.9862_{-0.0021}^{+0.0007}$ & $0.92_{-0.22}^{+0.07}$ \\
$\Gamma$ & $2.32_{-0.03}^{+0.05}$ & $2.12_{-0.03}^{+0.06}$ & $2.16_{-0.03}^{+0.04}$ &  $2.208_{-0.031}^{+0.016}$ & $1.978^\star$ \\
$\log\xi$ & $2.58_{-0.04}^{+0.08}$ & $3.02_{-0.09}^{+0.03}$ & $2.89_{-0.06}^{+0.04}$ & $2.82_{-0.04}^{+0.03}$ & $3.434^\star$ \\
$A_{\rm Fe}$ & $0.68_{-0.07}^{+0.06}$ & $0.50_{}^{+0.04}$ & $0.50_{}^{+0.05}$ & $0.50_{}^{+0.04}$ & $0.5^\star$ \\
$E_{\rm cut}$ [keV] & $88_{-3}^{+5}$ & $63.3_{-2.2}^{+1.5}$ & $68_{-3}^{+3}$ & $76.1_{-10.1}^{+1.7}$ & $50.9^\star$ \\
$R_\text{f}$ & $0.58_{-0.11}^{+0.07}$ & $0.43_{-0.07}^{+0.05}$ & $0.45_{-0.03}^{+0.10}$ & $0.52_{-0.04}^{+0.03}$ & $0.61^{+0.16}_{-0.15}$ \\
$\alpha$ & $-0.069_{-0.037}^{+0.014}$ & $0.076_{-0.018}^{+0.014}$ & $0.012_{-0.025}^{+0.017}$ & $-0.034_{-0.018}^{+0.019}$ & $0.04_{-0.03}^{+0.05}$ \\ 
\hline
\hline
$\chi^2$/dof & 2272.41/2208 & 2334.70/2209 & 2320.45/2209 & 2302.03/2209 & 2405.10/2213 \\
& =1.02917 & =1.05690 & =1.05045 & =1.04211 & = 1.08681 \\
\hline\hline
\end{tabular}
}
%\vspace{0.2cm}
\caption{Summary of the best-fit values for Model~1 ($i$ free), Model~2$a$ ($i=60^\circ$), Model~2$b$ ($i=65^\circ$), Model~2$c$ ($i=70^\circ$), and Model~3 (frozen $N_{\rm H}$, $\Gamma$, $\xi$, $A_{\rm Fe}$, and $E_{\rm cut}$). The reported uncertainties correspond to 90\% confidence level for one relevant parameter ($\Delta\chi^2 = 2.71$). $\xi$ in units of erg~cm~s$^{-1}$. $^\star$ indicates that the parameter is frozen in the fit. When the upper or lower uncertainties are not reported, the best-fit value is stuck at a boundary of the allowed range. $q_{\rm in}$ and $q_{\rm out}$ are allowed to vary in the range $[0,10]$ and $A_{\rm Fe}$ in the range $[0.5,10]$. \label{t-fit}}
\end{table*}

\begin{figure*}[t]
\begin{center}
\includegraphics[width=17.5cm]{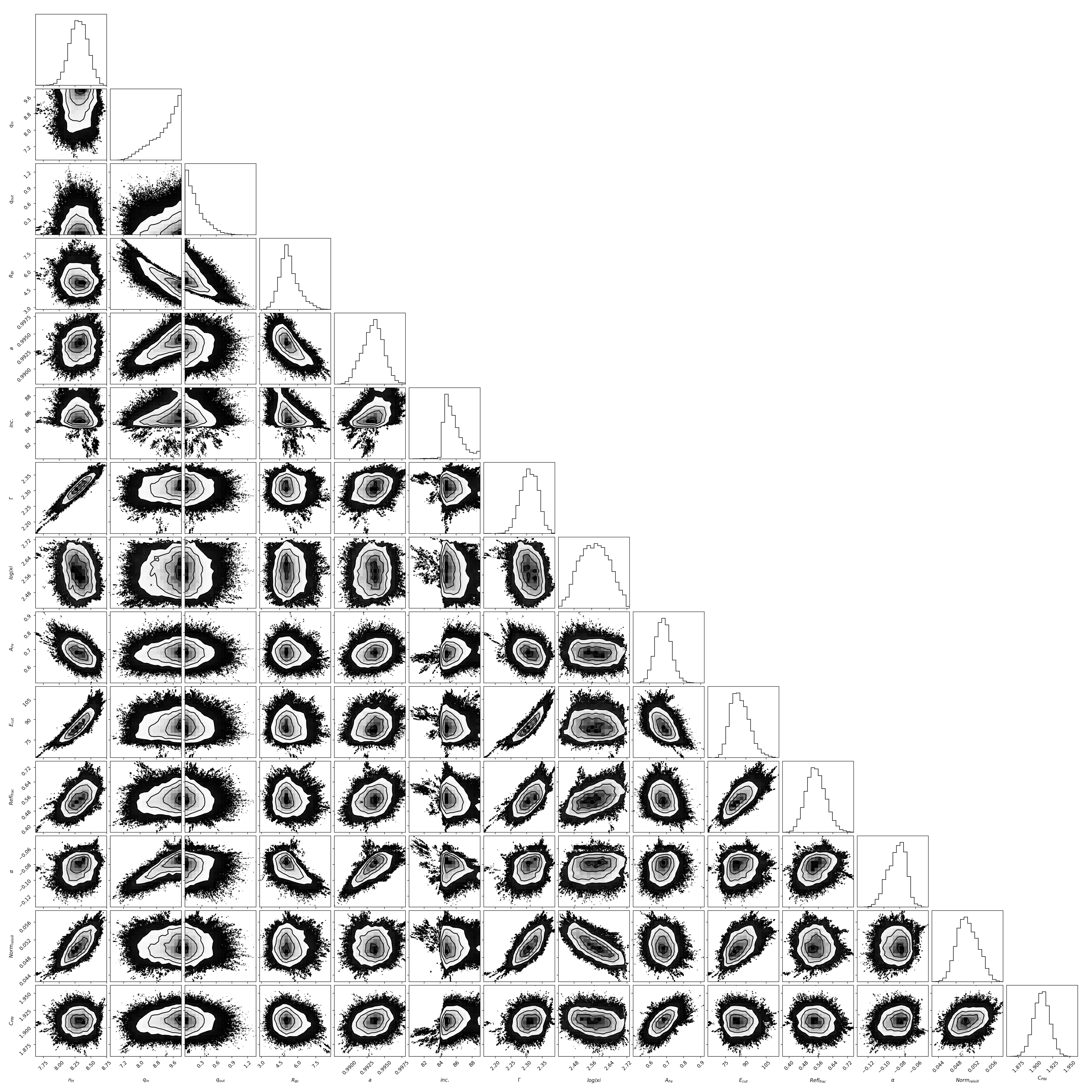}
\end{center}
\vspace{-0.5cm}
\caption{Corner plot for all the free parameter pairs for Model~1 with free inclination angle after the MCMC run. The contour levels in the 2D plots correspond to the 1-, 2-, and 3-$\sigma$ confidence contours. \label{f-mcmc}}
\end{figure*}

The results of the fit are shown in Tab.~\ref{t-fit} under the name Model~1. The model and the data to best-fit model ratio are shown in Fig.~\ref{f-mr}. Since we have thirteen free parameters in the model and the $\chi^2$ landscape may be complicated, we run a Markov Chain Monte Carlo (MCMC) analysis using the python script by Jeremy Sanders\footnote{Available on github at\\ \href{https://github.com/jeremysanders/xspec_emcee}{https://github.com/jeremysanders/xspec\_emcee}.}. We use 200~walkers of 10,000~iterations, after burning the initial 1,000~iterations (corresponding to $\sim 100$~times the autocorrelation length). The corner plot with the 1- and 2-dimensional projections of the posterior probability distributions is shown in Fig.~\ref{f-mcmc}. The results from the MCMC analysis agree well with the results from XSPEC. The discussion of our results is postponed to the next section.

\begin{figure}[t]
\begin{center}
\includegraphics[width=8.0cm]{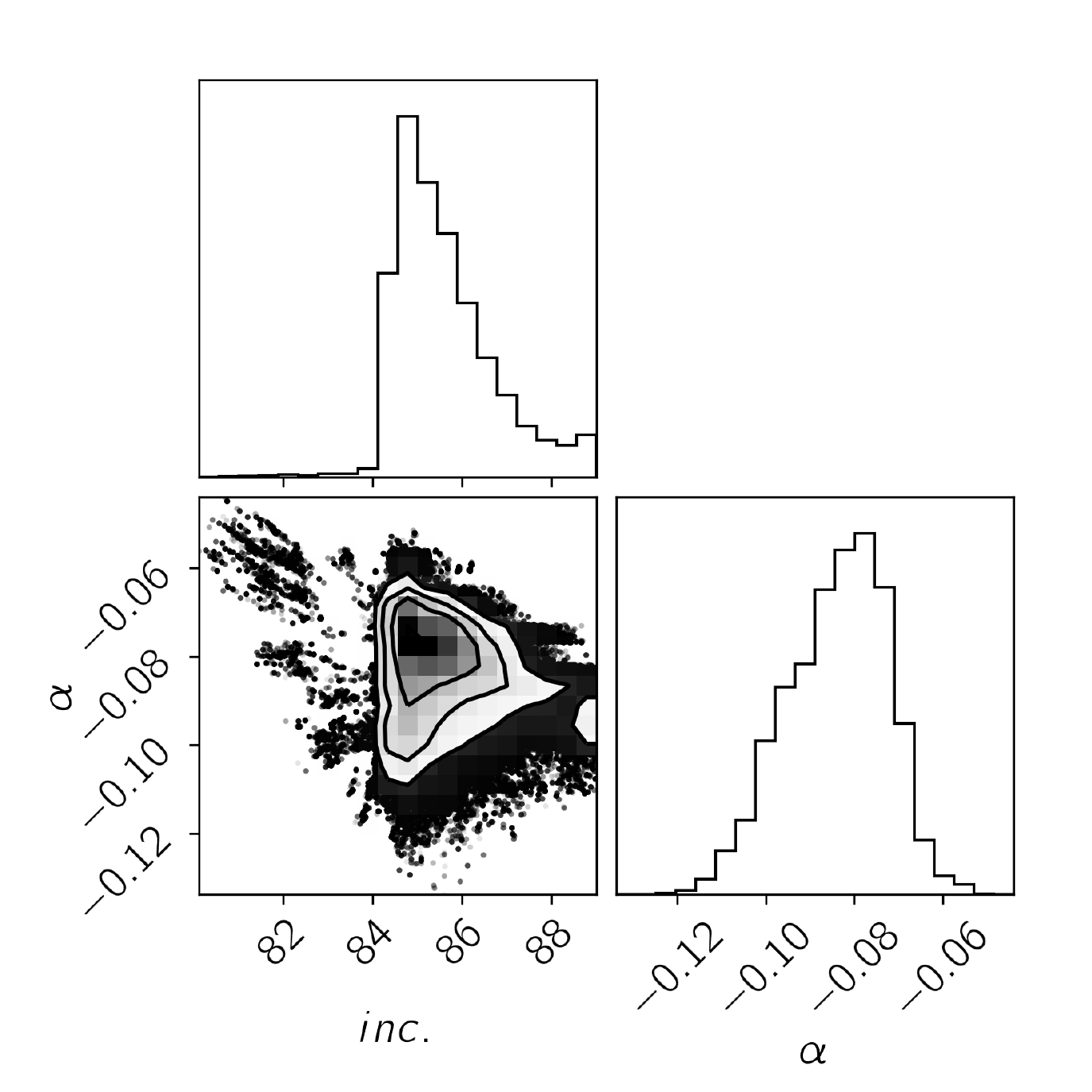}
\end{center}
\vspace{-0.6cm}
\caption{1-, 2-, and 3-$\sigma$ confidence contours for the inclination angle of the disk and the parameter $\alpha$ from Fig.~\ref{f-mcmc}. \label{f-mcmcz}}
\end{figure}

\section{Discussion and conclusions \label{s-dis}}

From the best-fit values for Model~1 in Tab.~\ref{t-fit}, we can immediately notice an exceptionally high viewing angle $i \approx 85^\circ$ and a negative value of $\alpha$ at a high confidence level; that is, the accretion disk around GRS~1915+105 would be a sub-Keplerian disk. However, such a high value of the viewing angle is quite suspicious and previous studies have reported values ranging from $\sim 55^\circ$ to $\sim 75^\circ$~\cite{fender,blum09,miller13,yuexin,tlb}. In Fig.~\ref{f-mcmcz}, we zoom the panel viewing angle $i$ vs parameter $\alpha$ of the corner plot in Fig.~\ref{f-mcmc} and we do not see any evidence of a correlation between $i$ and $\alpha$. However, the best-fit values in Tab.~\ref{t-fit} suggest a compensation between these two parameters. If the angular velocity of the disk is lower than the Keplerian angular velocity $\Omega_{\rm K}$, the effect of Doppler boosting is weaker. Such an effect can be compensated by increasing the viewing angle $i$, as the effect of Doppler boosting increases as we increase the value of $i$.

In order to investigate more the correlation between $i$ and $\alpha$, we consider three more models in which the inclination angle is frozen to $60^\circ$ (Model~2$a$), $65^\circ$ (Model~2$b$), and $70^\circ$ (Model~2$c$). The results of the fits are shown in Tab.~\ref{t-fit}. We can see that if we impose a lower viewing angle $i$, the value of $\alpha$ increases, which is consistent with our previous consideration that both parameters contribute to the Doppler boosting and therefore they can compensate each other. It is worth nothing that the most popular measurement of the viewing angle of the disk of the black hole in GRS~1915+105 is $i_{\rm jet} = 66^\circ \pm 2^\circ$~\cite{fender}, which is inferred from the measurement of the direction of the jet and assuming that the latter is parallel to the black hole spin and perpendicular to the accretion disk. If we believe in such a measurement of the viewing angle and and we freeze $i$ to that value in the fit, we find that the measurement of the parameter $\alpha$ is consistent with a Keplerian disk.

For other considerations on this observation and the choice of the model, the reader can find more details in previous studies~\cite{nk3,yuexin,tlb}. The data to best-fit model ratio in Fig.~\ref{f-mr} shows some unresolved features in the range 6-8~keV, but currently there is no clear explanation. If we add a distant reflector in the model, the fit improves only marginally: the minimum of $\chi^2$ is only a bit lower and the unresolved features remains there. So a distant reflector does not seem to be required. The emissivity profile with a very high $q_{\rm in}$ and an almost vanishing $q_{\rm out}$ was also found in previous analysis of this \textsl{Suzaku} observations~\cite{blum09,yuexin,tlb} as well as in a \textsl{NuSTAR} observation of this source~\cite{miller13} and of GS~1354--645~\cite{elbatal16,yerong}. Such a density profile may be explained with a ring-like or disk-like corona~\cite{miniutti03,wf11,wg15}. The very low iron abundance reported in Tab.~\ref{t-fit} might be due to the fact that {\sc relxill} neglects in its calculations the returning radiation, namely the radiation emitted by the disk and returning to the disk because of the strong light bending near the black hole, and the fact that the black hole spin parameter $a_*$ and the viewing angle $i$ are both high~\cite{shafqat3}.

\begin{figure*}[t]
\begin{center}
\includegraphics[width=14.0cm]{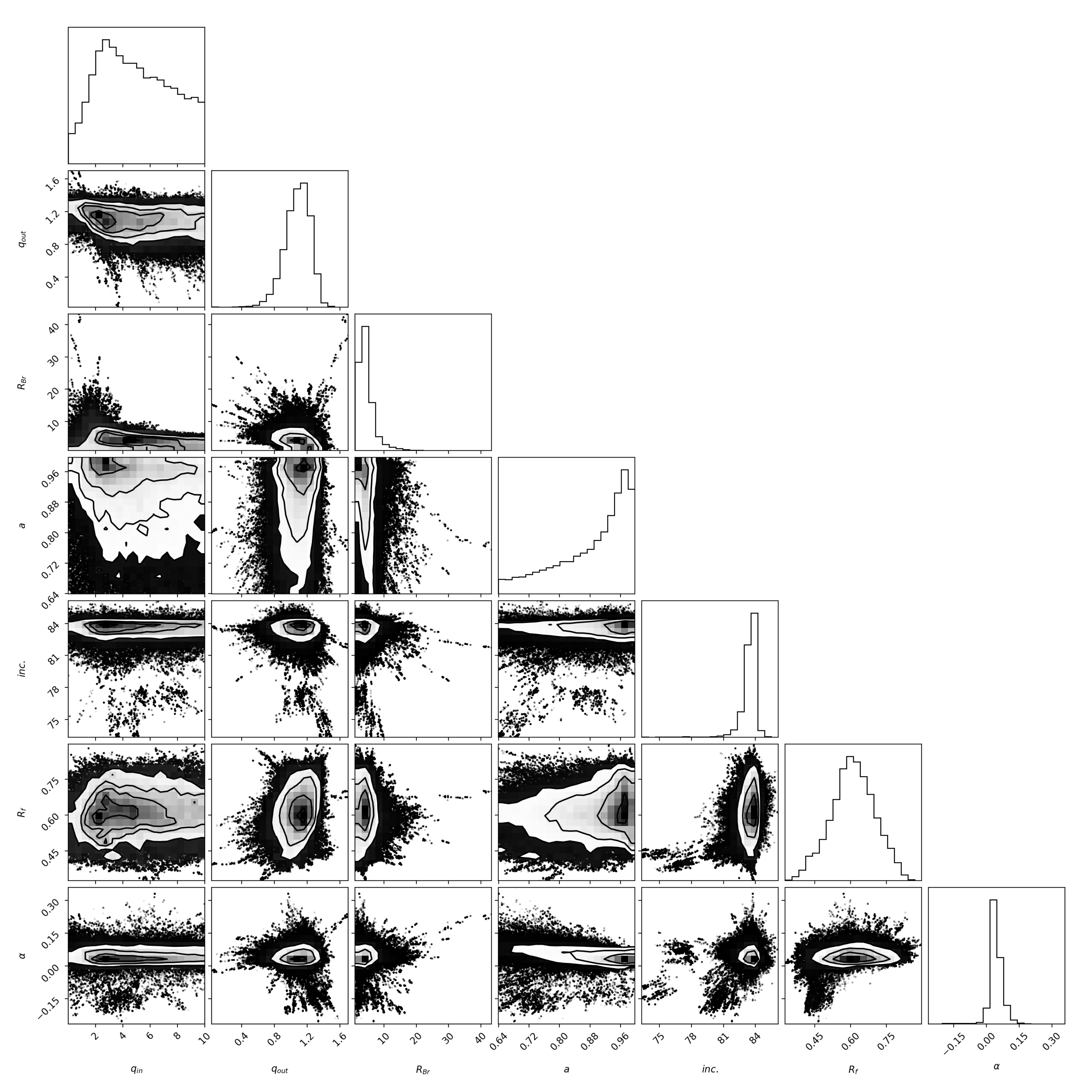}
\end{center}
\vspace{-0.5cm}
\caption{Corner plot for all the free parameter pairs for Model~3 with frozen $N_{\rm H}$, $\Gamma$, $\xi$, $A_{\rm Fe}$, and $E_{\rm cut}$, after the MCMC run. The contour levels in the 2D plots correspond to the 1-, 2-, and 3-$\sigma$ confidence contours. \label{f-mcmc3}}
\end{figure*}

In the effort to decrease the number of free parameters in the fit, with the goal of focusing our study to the parameter $\alpha$, we can proceed as follows. First, we fit the data removing the 5-9~keV energy band of the iron line, which is the most sensitive region to relativistic effects. In this fit, we freeze the parameters that are normally determined by the shape of the iron line and that would be unconstrained without the iron line region: we set $q_{\rm in} = q_{\rm out} = 3$, $a_* = 0.998$, $i = 66^\circ$, and $\alpha = 0$. For a different choice of the value of these parameters, our fit does not change significantly. After that, we run an MCMC analysis of the full spectrum, freezing the column density $N_{\rm H}$, the photon index $\Gamma$, the ionization parameter $\xi$, the iron abundance $A_{\rm Fe}$, and the high-energy cut-off $E_{\rm cut}$ to the values obtained in the previous fit without the iron line region. Now we have seven free parameters ($q_{\rm in}$, $q_{\rm out}$, $R_{\rm br}$, $i$, $a_*$, $\alpha$, and $R_{\rm f}$) that are instead sensitive to the shape of the iron line. This is our Model~3 and the best-fit values are shown in the last column in Tab.~\ref{t-fit}. The corner plot is shown in Fig.~\ref{f-mcmc3}. The fit is clearly worse and the uncertainties are larger, because we have a more constrained model. Unfortunately, the estimate of the parameter $\alpha$ is still strongly biased by the estimate of the viewing angle $i$, confirming our previous conclusion that X-ray reflection spectroscopy can potentially test the angular velocity of the accretion disk, but it would be essential to have an independent estimate of the viewing angle of the disk. Both parameters contribute to the strength of the Doppler boosting and they can thus compensate each other.

In conclusion, here we have proposed to test the angular velocity of the accretion disk in black hole binaries and AGNs using a modified version of the relativistic reflection model {\sc relxill}. We have applied the model to a \textsl{Suzaku} observation of the black hole binary GRS~1915+105 to explore the feasibility of our proposal. Our results suggest that we can test the Keplerian disk hypothesis in the presence of an independent and robust measurement of the viewing angle $i$. It is worth noting that, if we assume the viewing angle $i \approx 65^\circ$ inferred from the direction of the jet~\cite{fender}, which is currently the most accredited value in literature for GRS~1915+105, our results are consistent with the Keplerian disk hypothesis for this source.

%%%%%%%%%%%%%%%%%%%%%%%%%%%%%%%

\vspace{0.5cm}

{\bf Acknowledgments --}
This work was supported by the Innovation Program of the Shanghai Municipal Education Commission, Grant No.~2019-01-07-00-07-E00035, and the National Natural Science Foundation of China (NSFC), Grant No.~11973019.
A.T., C.B., and S.N. are members of the International Team~458 at the International Space Science Institute (ISSI), Bern, Switzerland, and acknowledge support from ISSI during the meetings in Bern.

%%%%%%%%%%%%%%%%%%%%%%%%%%%%%%%

\end{document}